White Paper for Plasma 2020 Decadal Survey

# Major Scientific Challenges and Opportunities in Understanding Magnetic Reconnection and Related Explosive Phenomena in Magnetized Plasmas

Hantao Ji
(Tel: 609-243-2162, Email: hji@princeton.edu)
*Princeton University*

February 2019

Co-authors:
A. Alt[1], S. Antiochos[2], S. Baalrud[3], S. Bale[4], P. M. Bellan[5], M. Begelman[6], A. Beresnyak[7], E.G. Blackman[8], D. Brennan[1], M. Brown[9], J. Buechner[10], J. Burch[11], P. Cassak[12], L.-J. Chen[2], Y. Chen[6], A. Chien[1], D. Craig[13], J. Dahlin[14], W. Daughton[15], E. DeLuca[16], C. F. Dong[1], S. Dorfman[17], J. Drake[18], F. Ebrahimi[19], J. Egedal[20], R. Ergun[6], G. Eyink[21], Y. Fan[22], G. Fiksel[23], C. Forest[20], W. Fox[19], D. Froula[8], K. Fujimoto[24], L. Gao[19], K. Genestreti[25], S. Gibson[22], M. Goldstein[26], F. Guo[15], M. Hesse[27], M. Hoshino[28], Q. Hu[29], Y.-M. Huang[1], J. Jara-Almonte[19], H. Karimabadi[30], J. Klimchuk[2], M. Kunz[1], K. Kusano[31], A. Lazarian[20], A. Le[15], H. Li[15], X. Li[15], Y. Lin[32], M. Linton[7], Y.-H. Liu[33], W. Liu[34], D. Longcope[35], N. Loureiro[36], Q.-M. Lu[37], Z-W. Ma[38], W. H. Matthaeus[39], D. Meyerhofer[15], F. Mozer[4], T. Munsat[6], N. A. Murphy[16], P. Nilson[8], Y. Ono[28], M. Opher[40], H. Park[41], S. Parker[6], M. Petropoulou[1], T. Phan[4], S. Prager[1], M. Rempel[22], C. Ren[8], Y. Ren[19], R. Rosner[42], V. Roytershteyn[17], J. Sarff[20], A. Savcheva[16], D. Schaffner[43], K. Schoeffier[44], E. Scime[12], M. Shay[39], M. Sitnov[45], A. Stanier[15], J. TenBarge[1], T. Tharp[46], D. Uzdensky[6], A. Vaivads[47], M. Velli[48], E. Vishniac[21], H. Wang[49], G. Werner[6], C. Xiao[50], M. Yamada[19], T. Yokoyama[28], J. Yoo[19], S. Zenitani[51], E. Zweibel[20]

*[1]Princeton U., [2]NASA GSFC, [3]U. Iowa, [4]UC Berkeley, [5]Caltech, [6]U. Colorado – Boulder, [7]NRL, [8]U. Rochester, [9]Swarthmore College, [10]Max Planck for Solar System Research, Germany, [11]Southwest Research Institute, [12]West Virginia U., [13]Wheaton College, [14]Universities Space Research Association (USRA), [15]LANL, [16]Harvard Smithsonian Center for Astrophysics, [17]Space Science Institute, [18]U. Maryland, [19]PPPL, [20]U. Wisconsin – Madison, [21]Johns Hopkins U., [22]High Altitude Observatory, [23]U. Michigan, [24]Beihang U., China, [25]U. New Hampshire, [26]Univ. Maryland Balt. County, [27]U. Bergen, Norway, [28]U. Tokyo, Japan, [29]U. Alabama – Huntsville, [30]Analytics Ventures, [31]Nagoya U., Japan, [32]Auburn U., [33]Dartmouth College, [34]Lockheed Martin Solar & Astro Lab (LMSAL), [35]Montana State U., [36]MIT, [37]U. Science and Technology of China, [38]Zhejiang U., China, [39]U. Delaware, [40]Boston U., [41]UNIST, Korea, [42]U Chicago, [43]Bryn Mawr College, [44]Instituto Superior Technico, Portugal, [45]APL, [46]Marquette U., [47]Swedish Institute of Space Physics, [48]UCLA, [49]New Jersey Inst. Tech., [50]Peking U., China, [51]Kyoto U., Japan*

(100 authors and 51 institutions)

## Major Scientific Challenges and Opportunities in Understanding Magnetic Reconnection and Related Explosive Phenomena in Magnetized Plasmas

### I. Magnetic Reconnection: A Fundamental Process throughout the Universe and in the Lab

Magnetic reconnection - the topological rearrangement of magnetic field - underlies many explosive phenomena across a wide range of natural and laboratory plasmas [1-3]. It plays a pivotal role in electron and ion heating, particle acceleration to high energies, energy transport, and self-organization. Reconnection can have a complex relationship with turbulence at both large and small scales, leading to various effects which are only beginning to be understood. **In heliophysics**, magnetic reconnection plays a key role in solar flares, coronal mass ejections, coronal heating, solar wind dissipation, the interaction of interplanetary plasma with magnetospheres, dynamics of planetary magnetospheres such as magnetic substorms, and the heliospheric boundary with the interstellar medium. Magnetic reconnection is integral to the solar and planetary dynamo processes. **In astrophysics**, magnetic reconnection is an important aspect of star formation in molecular clouds, stellar flares, explosive phenomena from magnetars and pulsars (including Crab Nebula), and even for acceleration of cosmic rays at ultra-high energies. Magnetic reconnection is thought to occur in both coronae and interiors of magnetized accretion disks in proto-stellar systems and X-ray binaries, as well as in interstellar medium turbulence. Magnetic reconnection is believed to occur in the centers of Active Galactic Nuclei, where matter is accreted onto supermassive black holes. On even larger scales, magnetic reconnection may be important in extragalactic radio jets and lobes, and even in galaxy clusters. Magnetic reconnection might occur during the recently discovered Fast Radio Bursts. **In laboratory plasmas**, magnetic reconnection is known to occur during sawtooth oscillations in tokamaks, neoclassical tearing mode growth, disruptions, the startup of Spherical Torus plasmas using Coaxial Helicity Injection, relaxation in Reversed Field Pinches and spheromaks, the formation of Field Reversed Configurations by theta pinch or plasma merging, and possibly in edge-localized modes. Magnetic reconnection may play a role in magnetized inertial fusion plasmas such as Z pinches or laser plasmas. Thus, understanding magnetic reconnection is of fundamental importance for plasma physics and significantly contributes to our understanding of the Universe and to the success of fusion energy.

### II. Major Scientific Challenges in Understanding Reconnection and Related Explosive Phenomena

Despite significant progress in the past decades, the following major scientific challenges must be resolved before reconnection is fully understood.

**1. The multiple scale problem** [e.g., 3-19]: Reconnection involves the coupling between the large MHD scale of the system and the kinetic ion and electron dissipation scales that are orders of magnitude smaller. This coupling is currently not well understood, and the lack of proper fluid closure models is the core of the multiple scale problem. The phase diagram illustrated in Fig. 1 is based on plasmoid dynamics and shows how different coupling mechanism can exist. Questions here include: how do plasmoid dynamics scale with key parameters, such as the Lundquist number, how is this scaling influenced by a guide field, do there exist other coupling mechanisms, and how does reconnection respond to turbulence and associated dissipation on scales below or above the electron scales.

**2. The 3D Problem** [e.g., 15-23]: Numerous studies have focused on reconnection in 2D while natural plasmas are 3D. It is critical to understand which features of 2D systems carry over to 3D and which are fundamentally altered. New effects that require topological analysis include instabilities due to variations in the third direction leading to complex interacting “flux ropes”, enhanced magnetic stochasticity, and field line separation in 3D. How fast reconnection is related to self-organization phenomena, such as Taylor relaxation, as well as the accumulation of magnetic helicity remains a longstanding problem.

**3. Energy conversion** [e.g., 24-35]: Reconnection is invoked to explain the observed conversion of magnetic energy to heat or to non-thermal particle energy. A major challenge in connecting theories and experiments to observations is the ability to quantify the detailed energy conversion process. There exist competing theories of particle acceleration based on 2D and 3D reconnection, but as of yet there is no consensus on the origin of the observed power laws in particle energy distributions.

**4. Boundary conditions** [e.g., 36-39]: It is unclear that an understanding of reconnection physics in periodic systems can be directly applied to natural plasmas which are non-periodic, and often line-tied at their ends such as in solar flares. Whether line-tying and driving from the boundaries fundamentally alters reconnection physics has profound importance in connecting laboratory physics to astrophysics. It is also important to understand how reconnection works in naturally occurring settings which have background flows, out-of-plane magnetic fields, and asymmetries.

**5. Onset** [e.g., 40-45]: Reconnection in natural and laboratory plasmas often occurs impulsively, with a rapid onset, following a slow energy building up process. Is the onset a local and spontaneous (e.g., plasmoid instability) or a globally driven process (e.g., ideal MHD instabilities), and is the onset mechanism a 2D or 3D phenomenon? What are the roles of collisions or global magnetic geometry on the onset conditions? A related important question is how magnetic energy is accumulated and stored prior to onset.

**6. Partial ionization** [e.g., 46-50]: Important reconnection events often occur in weakly ionized plasmas, such as solar chromosphere, whose heating requirements dwarf those of corona. This introduces new physics associated with neutral particles. Questions include whether increased friction slows down reconnection or enhanced two-fluid effects accelerates reconnection.

**7. Flow-driven** [e.g., 44,51]: Magnetic field is generated by dynamo mechanisms in flow-driven systems such as stars and accretion disks, and reconnection is thought to be an integral part of the dynamo process. Key questions include under what conditions reconnection can occur in such systems? How fast does it proceed? What effect does reconnection have on the associated turbulence and transport?

**8. Extreme conditions** [e.g., 52]: Extreme plasma conditions in the vicinity of compact astrophysical objects (e.g., intense radiation and pair-particle creation) must be included in analytical and numerical models of reconnection before applying theory to observations.

**9. Turbulence, shock and reconnection** [e.g., 14-17,53-60] Reconnection is closely interconnected to other fundamental plasma processes such as turbulence and shocks. It is important to understand the rates of topology change, energy release, and heating, as they may be tied to the overall turbulence and shock dynamics.

**10. Related explosive phenomena** [e.g., 36,41,61] Global ideal MHD instabilities, both linear (kink, torus) and nonlinear (ballooning), are closely related to magnetic reconnection either as a driver or a consequence such as in Coronal Mass Ejections, magnetic storms/substorms, and dipolarization fronts in the magnetotail. Understanding how, and under what conditions, such explosive phenomena take place, as well as their impact, remain major scientific challenges.

## III. Major Research Opportunities in Meeting Scientific Challenges

The table below summarizes the capabilities to address the major reconnection questions listed in Sec.II by each approach: theory/simulation, observations (in situ and remote sensing), and laboratory experiments (magnetized basic, magnetic fusion, and high energy density experiments). Green indicates that required capabilities are already in place or can rapidly be made available given adequate investment from funding agencies; yellow indicates substantial difficulties which are unlikely to be resolved within the next decade.

| Questions | Theory & Simulation | Observations | | Laboratory Experiments | | |
|---|---|---|---|---|---|---|
| | | *in situ* | Remote Sensing | Magnetized Basic | Magnetic Fusion | High Energy Density |
| Multi-Scale | [green] | [green] | [yellow] | [green] | [green] | [green] |
| 3D | [green] | [green] | [green] | [green] | [green] | [green] |
| Energy | [green] | [green] | [green] | [green] | [green] | [green] |
| Boundary | [green] | [yellow] | [green] | [green] | [yellow] | [yellow] |
| Onset | [green] | [green] | [green] | [green] | [green] | [yellow] |
| Partial Ionization | [green] | [yellow] | [green] | [green] | [yellow] | [yellow] |
| Flow-driven | [green] | [green] | [green] | [green] | [yellow] | [green] |
| Extreme | [green] | [yellow] | [green] | [yellow] | [yellow] | [green] |
| Turbulence/shock | [green] | [green] | [green] | [green] | [yellow] | [yellow] |
| Explosive | [green] | [green] | [green] | [green] | [green] | [green] |

**1. Theory and numerical simulation**. These major questions are ripe for study with fluid, kinetic, and hybrid models, as well as, multi-fluid models that include kinetic effects through physics-based closure equations. Exascale computing will allow us to finally attack these questions in fully 3D systems with realistic plasma parameters. Dedicated programs from the DoE and NSF are needed that, over the next decade, will support the development of next-generation analytical and numerical models and the application of new numerical technology and theoretical understanding to the reconnection challenges.

**2. Observations**. Over the next decade unprecedented remote sensing observations from missions such as Solar Dynamics Observatory, Interface Region Imaging Spectrograph, Fermi, and the upcoming Solar Orbiter, as well as exquisite in situ measurements from missions such as Magnetospheric Multiscale, Parker Solar Probe and BepiColombo, will deliver a wealth of critical new insights on reconnection. The modeling and theory programs proposed above will be ideally positioned to take advantage of these data for advancing and validating understanding.

**3. Laboratory experiments**. Ground-truth in new understanding is ultimately provided by comparing theory and models with laboratory experiments. Dedicated basic reconnection experiments have greatly matured, and next-generation facilities are poised to provide access to new reconnection phases with direct relevance to heliophysics and astrophysics. New diagnostics are needed to provide kinetic-scale measurements at the distribution function level. Magnetic fusion and high-energy-density experiments provide unique platforms to address certain major questions. In concert with theory and simulation, dedicated programs from the DoE and NSF are urgently needed to support the next-generation experimental facilities and diagnostic instrumentation that are absolutely essential for achieving closure between theory and experiment.

In summary, magnetic reconnection is a fundamental process both in the universe and in laboratory plasmas. The new research capabilities in theory/simulations, observations, and laboratory experiments provide the opportunity to finally solve the grand scientific challenges summarized in this whitepaper. Success will require enhanced and sustained investments from relevant funding agencies (DoE and NSF) throughout the coming decade, especially in the areas of theory/simulations and laboratory experiments covered specifically by this survey. These investments will deliver transformative progress in understanding magnetic reconnection, which will benefit many areas of critical practical importance, such as fusion and space weather.

**References**

[1] E. Zweibel and M. Yamada, Ann. Rev. Astron. Astrophys. **47**, 291 (2009).
[2] M. Yamada, R. Kulsrud, and H. Ji, Rev. Mod. Phy. **82**, 603 (2010).
[3] H. Ji and W. Daughton, Phys. Plasmas **18**, 111207 (2011).
[4] N. Loureiro, A. Schekochihin, and S. Cowley, Phys. Plasmas **14**, 100703 (2007).
[5] W Daughton, V Roytershteyn, B. Albright et al., Phys. Rev. Lett. **103**, 065004 (2009).
[6] A. Bhattacharjee, Y.-M Huang, H. Yang, and B. Rogers, Phys. Plasmas **16**, 112102 (2009).
[7] A. Le, J. Egedal, W. Daughton et al., J. Plasma Phys. **81**, 305810108 (2014).
[8] J. Birn et al., J. Geophys. Res. **106**, 3715 (2001).
[9] M. Hesse, J. Birn, and M. Kuznetsova, J. Geophys. Res. **106**, 3721 (2001).
[10] Y. Ren, M. Yamada, S. Gerhardt, H. Ji et al., Phys. Rev. Lett. **95**, 055003 (2005).
[11] D. Uzdensky, N. Loureiro, and A. Schekochihin, Phys. Rev. Lett. **105**, 235002 (2010).
[12] A. L. Moser and P. M. Bellan, Nature **482**, 379 (2012).
[13] W. H. Matthaeus and M. Velli, Space Sci. Rev. **160**, 145 (2011).
[14] W. Matthaeus and S. Lamkin, Phys. Fluids **28**, 303 (1985).
[15] A. Lazarian and E. Vishniac, Astrophys. J. **517**, 700 (1999).
[16] W. Daughton, V. Roytershteyn, H. Karimabadi et al., Nature Phys. **7**, 539 (2011).
[17] G. Eyink, A. Lazarian, and E. Vishniac, Astrophys. J. **743**, 51 (2011).
[18] J. Jara-Almonte, W. Daughton, and H. Ji, Phys. Plasmas **21**, 032114 (2014)
[19] F. Ebrahimi and R. Raman, Phys. Rev. Lett. **114**, 205003 (2015).
[20] J.B. Taylor, Rev. Mod. Phys. **58**, 741 (1986).
[21] M. Zhang, N. Flyer, and BC Low, Astrophys. J. **644**, 575 (2006).
[22] F. Ebrahimi, Phys. Plasmas **23**, 120705 (2016).
[23] A. Boozer, Phys. Plasmas **19**, 092902 (2012).
[24] S. Krucker, H. Hudson, L. Glesener et al., Astrophys. J. **714**, 1108(2010).
[25] E. Scime, S. Hokin, N. Mattor, and C. Watts, Phys. Rev. Lett. **68**, 2165(1992).
[26] Y. Ono, M. Yamada, T. Akao et al., Phys. Rev. Lett. **76**, 3328(1996).
[27] M. Brown, C. Cothran, M. Landreman et al., Astrophys. J. **577**, L63 (2002).
[28] J. Egedal, W. Daughton, and A. Le, Nature Phys. **8**, 321 (2012).
[29] J. Drake, M. Swisdak, H. Che, and M. A. Shay, Nature **443**, 553 (2006).
[30] M. Hoshino, Phys. Rev. Lett. **108**, 135003 (2012).
[31] M. Yamada, J. Yoo, J. Jara-Almonte, H. Ji et al., Nature Communications **5**, 4774 (2014).
[32] G. Werner et al., Mon. Not. R. Astron. Soc. **473**, 4840 (2018).
[33] Y. D. Yoon and P. M. Bellan, Astrophys. J. Lett. **868**, L31 (2018).
[34] R. S. Marshall, M. J. Flynn, and P. M. Bellan, Phys. Plasmas **25**, 11210 (2018)
[35] M. Sitnov, V. Merkin, V. Roytershteyn, M. Swisdak, Geophys. Res. Lett. **45**, 4639 (2018).
[36] K. Shibata, S. Masuda, M. Shimojo, H. Hara et al., Astrophys. J. **451**, L83 (1995).
[37] W. Bergerson, C. Forest, G. Fiksel, D. Hannum et al., Phys. Rev. Lett. **96**, 015004 (2006).
[38] I. Furno, T. Intrator, G. Lapenta, L. Dorf et al., Phys. Plasmas **14**, 022103 (2007).
[39] C. Myers, M. Yamada, H. Ji et al., Nature **528**, 526 (2015).
[40] P. Cassak, M. Shay, and J. Drake. Phys. Rev. Lett. **95**, 235002 (2005).
[41] J. Klimchuk, Phil. Trans. R. Soc. A **373**, 20140256 (2015).
[42] N. Katz, J. Egedal, W. Fox, A. Le et al., Phys. Rev. Lett. **104**, 255004 (2010).
[43] F. Pucci and M. Velli, Astrophys. J. Lett. **780**, L19 (2014).
[44] P. Pritchett, Phys. Plasmas **22**, 062102 (2015).
[45] D. Uzdensky and N. Loureiro, Phys. Rev. Lett. **116**, 105003 (2016).
[46] A. Lazarian, E. Vishniac, and J. Cho, Astrophys. J. **603**, 180 (2004).
[47] E. Zweibel, E. Lawrence, J. Yoo, H. Ji et al., Phys. Plasmas **18**, 111211 (2011).
[48] J. Leake, V.S. Lukin, M. Linton, and E. Meier, Astrophys. J. **760**, 109 (2012).
[49] E. Lawrence, H. Ji, M. Yamada, and J. Yoo, Phys. Rev. Lett. **110**, 015001 (2013).

[50] J. Jara-Almonte, H. Ji et al., Phys. Rev. Lett. **122**, 015101 (2019).
[51] G. Fiksel, W. Fox, A. Bhattacharjee et al., Phys. Rev. Lett. **113**, 105003 (2014).
[52] D. Uzdensky, Space Sci. Rev. **160**, 45 (2011).
[53] S. Servidio, W. H. Matthaeus, M. A. Shay et al., Phys. Plasmas **17**, 032315 (2010).
[54] V. Zhdankin, D. Uzdensky, J. C. Perez amd S. Boldyrev, Astrophys. J. **771**, 124 (2013)
[55] H. Karimabadi et al., Phys. Plasmas **21**, 062308 (2014).
[56] Y. Matsumoto et al., Science **347**, 974 (2015).
[57] C. C. Haggerty, T. N. Parashar, W. H. Matthaeus et al., Phys Plasmas, **24**, 102308 (2017).
[58] N. Loureiro and S. Boldyrev, Phys. Rev. Lett. **118**, 2451010 (2017).
[59] A. Mallet, A. Schekochihin, and B. Chandran, Mon. Not. R. Astron. Soc. **468**, 4862 (2017).
[60] C. Dong, L. Wang, Y.-M. Huang et al., Phys. Rev. Lett. **121**, 165101 (2018).
[61] A. Runov, V. Angelopoulos, M. Sitnov et al., Geophys. Res. Lett. **36**, L14106 (2009).

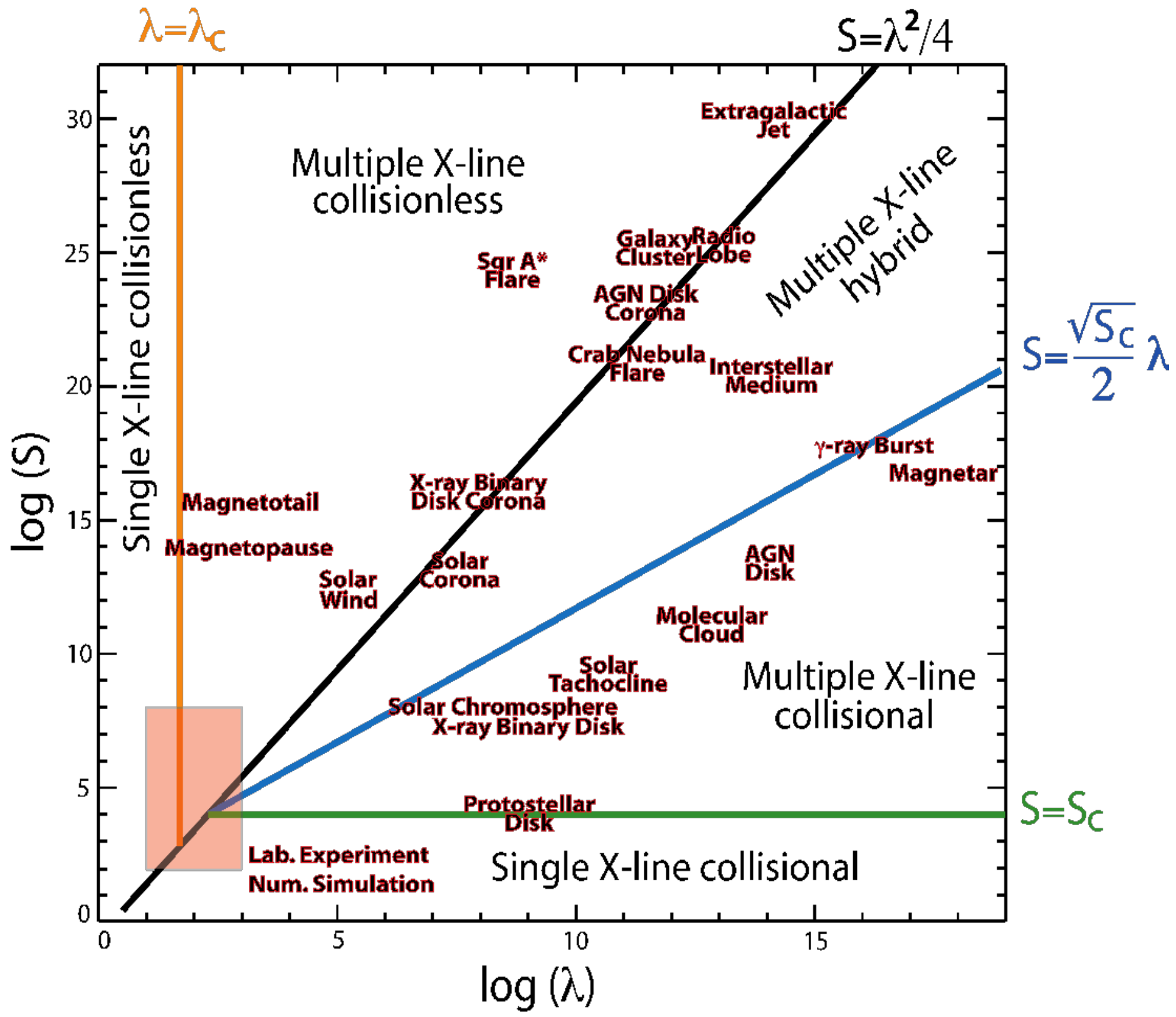

**Figure** 1. A phase diagram for magnetic reconnection. If either Lundquist number $S$ or the normalized size $\lambda$ (to ion kinetic scales), is small, reconnection with a single X-line occurs in collisional or in collisionless phases. When both $S$ and $\lambda$ are sufficiently large, three new multiple X-line phases appear through plasmoid instabilities. Global MHD physics can effectively couple to local dissipation scales, which can be either collisionless or collisional, depending on parameters. There are some updates on this diagram by including effects such as electron pressure anisotropy [7]. Various heliophysical and astrophysical cases are also shown in their approximate locations in the diagram. Numerical simulations and laboratory experiments can already, or are poised to, access all reconnection phases directly relevant to heliophysical and astrophysical plasmas. An particularly important question that should be addressed by these studies is how does this simplified 2D reconnection diagram extend into more realistic 3D. Figure adapted from Ref.[3]: H. Ji & W. Daughton, Phys. Plasmas **18**, 111207 (2011).